\documentclass[12pt]{article}
\usepackage{epsfig}
\begin{document}

\begin {center}
{\bf Review of Scalar Mesons}
\vskip 5mm
{D.\ V.\ Bugg\footnote{email: d.bugg@rl.ac.uk} \\
{\normalsize\it Queen Mary, University of London, London E1\,4NS, UK}
\\[3mm]}
\end {center}
\date{\today}

\begin{abstract}
\noindent
The formulae needed to parametrise $\sigma$, $\kappa$, $a_0(980)$
and $f_0(980)$ are reviewed.
For $\sigma$ and $\kappa$, the Adler zero due to Chiral Symmetry Breaking
plays a crucial role.
The $f_0(980)$ and $a_0(980)$ are locked to the $KK$ threshold by
a cusp mechanism which leads to a sharp peak in attraction at the
threshold.
This mechanism may play a wider role.
A novel suggestion is that the confinement potential involves
a cooperative effect between QCD effects (coloured quarks and gluons)
and conventional meson exchanges.

\vskip 2mm

\end{abstract}

\section {The $\sigma$ and $\kappa$ poles}
There is a long history to Chiral Symmetry Breaking.
It began with CVC and the Goldberger-Treiman relation \cite {Treiman},
and proceeded through the ground-breaking work of Gell-Mann and
L\' evy who invented PCAC and the linear and non-linear $\sigma$
models \cite {GMLevy}.
Their Lagrangian began from chiral symmetry in the sense that both
$\pi$ and $\sigma$ fields appear on an equal footing, but the
symmetry between them is spontaneously broken by introducing a
negative $\phi ^4$ term.

Adler \cite {Adler} showed that in both $\pi N$ and $\pi \pi$ systems
there is a zero at the unphysical point $t = m_\pi^2$ midway between the
thresholds of $s$ and $u$ channels.
`Soft' pion theory, current algebra and QCD led to the Standard Model.
With the idea that `bare' quarks are almost massless,
Gasser and Leutwyler introduced Chiral Perturbation Theory as a
systematic method of expanding amplitudes in the domain where pion
masses and momenta are small \cite {Gasser}.

The S-wave elastic scattering amplitude may be written
\begin {equation}
f_S(el) = N(s)/D(s) = K/(1 - iK\rho),
\end {equation}
where $\rho$ is the usual phase space factor.
The numerator $N(s)$ is real and describes `driving forces' from the
left-hand cut.
Unitarity is accomodated by means of the $K$-matrix.
The Adler zero in the $\pi \pi$ amplitude makes the S-wave unusual.
When one projects out the S-wave from the full amplitude, there is
a zero at $s = s_A \simeq m_\pi ^2/2$.
The amplitude near the $\pi \pi$ threshold rises nearly linearly.
The simplest acceptable form is \cite {Comments}
\begin {equation}
K_{\pi \pi } = b(s - s_A)\exp [-\gamma (s - M^2)],
\end {equation}
where $b$, $M$ and $\gamma$ are constants; the exponential makes
$K \to 0$ as $s \to \infty$ but introduces non-linearity in a
controlled way.
Above the $KK$ threshold
\begin {equation}
D(s) = M^2 - s - i[K_{\pi \pi}\rho_{\pi \pi } + g^2_{KK}\rho_{KK}].
\end {equation}

A similar approach has been used by Pelaez, Oset and Oller \cite {Pelaez}
\cite {spain} in a series of papers building Chiral Symmetry Breaking into
fits to data.
During the era 1975 to 1995, many theorists fitted the $\pi \pi$ S-wave
in ways similar to this and found a pole with mass roughly 500 MeV
and a large width $\sim$ 500 MeV.

In elastic scattering, the Adler zero in the numerator makes the
amplitude small near the $\pi \pi$ threshold.
However, in a production process, the numerator need not contain
the Adler zero, and is in fact consistent within rather small
errors with a constant.
The $\sigma$ pole appeared clearly in E791 data on
$D^+ \to \pi ^-\pi ^+ \pi ^+$ \cite {Aitala} and BES data on
$J/\Psi \to \omega \pi ^+ \pi ^-$ \cite {BESWPP}.
Both sets of data show a conspicuous peak visible by eye at
$\sim 500$ MeV.
Fig. 1 shows the Dalitz plot and mass projections for BES data.
There are conspicuous diagonal bands due to the $\sigma$ and
$f_2(1270)$ and vertical/horizontal bands due to
$b_1(1235) \to \omega \pi$.
The $\pi \pi$ mass projection is shown in (b) and the $\sigma$
component in (d).
Those data may be fitted simultaneously with elastic
scattering data; BES found a pole at $541 \pm 39 - i(252 \pm 42)$ MeV.

\begin {figure} [htb]
\begin {center}
\vskip -12mm
\epsfig {file=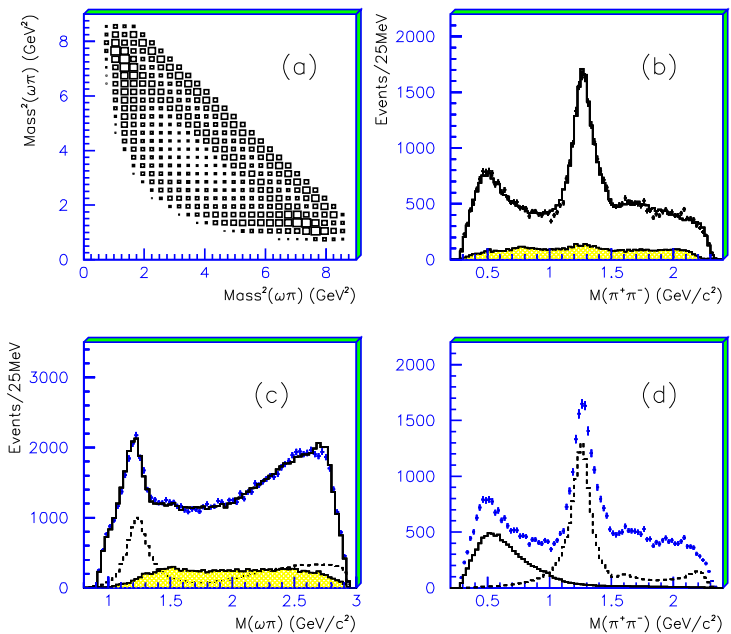,width=10cm}\
\vskip -51.5mm
\caption {BES data for $J/\Psi \to \omega \pi ^+ \pi ^-$.
(a) Dalitz plot; (b) $\pi \pi$ mass projection: the upper histogram
shows the fit, the lower one shows experimental
background;
(c) $\omega \pi$ mass projection; full histograms are as in (b) and
the dashed histogram shows the coherent sum of both $b_1(1235)\pi$
contributions; (d) $\pi \pi$ mass projections from $\sigma$ (full
curve) and spin 2 (dashed).}
\end {center}
\end {figure}

Since 2004, couplings to $KK$ and $\eta \eta$ have been determined by
fitting (a) all available data on $\pi \pi \to KK$ and $\eta \eta$,
(b) Kloe and Novosibirsk data on $\phi \to \gamma (\pi ^0 \pi ^0)$
\cite {Recon}.
All these data agree on a substantial coupling of $\sigma$ to $KK$.
A further refinement is to include in the Breit-Wigner denominator
a dispersive correction $m(s)$ (discussed below).
The $\sigma$ pole moves to $500 \pm 30 - i(264 \pm 30)$ MeV.

There was some debate whether the peak could be
due to interference with a `background'.
However, BES data give a phase variation with $s$ consistent with
elastic scattering and hence a simple pole \cite {sigphase}.
Secondly, Caprini et al. calculated a prediction  from
the Roy equations and driving forces on the left-hand cut \cite
{Caprini}.
The great merit of this calculation is that it maps out the
magnitude and phase of the S-wave amplitude over the whole of the
low mass region of the complex $s$-plane, including the sub-threshold
region inaccessible to experiment.
This leaves no possible doubt of the existence of the pole and gives a
pole position $M = 441^{+16}_{-8} -i(272 ^{+9}_{-12.5})$ MeV.
However, the result for $D(s)$ does not reproduce accurately the upper
side of the $\sigma$ peak in BES data.
A combined fit to their predictions and BES data using $(b_1 + b_2s)$
instead of $b$ in Eq. (2) gives $M = 472 \pm 30 - i(271 \pm 30)$ MeV
without departing from their predicted phases by more than $1.5^\circ$
\cite {sigpole}.

Understanding the $\sigma$ pole requires some gymnastics in complex
variables.
It lies at $s = 0.15 -i0.26$ GeV$^2$, not too far above the
$\pi \pi$ threshold but far into the complex plane.
The phase shift is $90^\circ$ just above the pole.
But on the real $s$-axis where experiments are done, the phase shift
must go to 0 at the $\pi \pi$ threshold.
There is therefore a strong variation of the phase as one moves into
the complex plane.
This arises directly from the Cauchy-Riemann equations: the Adler
zero gives a real part of the amplitude rising nearly linearly with
$s$ at low mass, and there must be a corresponding variation of the
imaginary part of the amplitude with ${\rm Im }\, s$.
Roughly speaking, the pole would give rise to a phase shift approaching
$150^\circ$ at 1 GeV, but the effect of unitarity is to distort the
phase observed on the real $s$-axis to $90^\circ$ at 1 GeV.
For the $\kappa$, the effect is even larger.
The $\kappa$ pole lies almost below the $K\pi$ threshold and is
very broad.
The consequence is that the phase shift measured on the real $s$ axis
only reaches $70^\circ$ at 1400 MeV/c.

At the Hadron95 conference, Mike Scadron asked me if there could be a
$\kappa$ resonance below 1 GeV.
At the time, the only reply which could
be given was that it must be very broad if it exists at all.
Later, Zheng and collaborators \cite {Zheng} showed that including the
Adler zero in the fit to LASS phase shifts for $K\pi$ elastic
scattering and including the effects of unitarity and analyticity
demands a pole at $M=694 \pm 53$ MeV, $\Gamma = 606 \pm 59$ MeV.
Descotes-Genon and Moussallam
obtained a pole at $M = 658 \pm 13$ MeV, $\Gamma = 557 \pm 24$ MeV
\cite {Descotes} using the Roy equations.

A low mass peak appears in E791 data on $D^+ \to K^- \pi ^+ \pi ^+$
\cite {E791} and BES data on $J/\Psi \to K^+\pi^-K^-\pi ^+$
\cite {BESkappa}.
Experimentalists have been bewildered by the strange behaviour of
the $K\pi$ phase shift and have undertaken the task of
determining the magnitude and phase of the $S$-wave amplitude in
bins of mass up to 1700 MeV/c \cite {Meadows}.
It turns out that these E791 results, the BES data and LASS phase
shifts may all be fitted simultaneously \cite {E791fit} with
$M = 750 ^{+30}_{-55} - i(342 \pm 60)$ MeV.
This result is rather higher in mass than the prediction from the
Roy equations, but again much of the discrepancy can be traced to
the effect of the strong $K\eta'$ threshold, which was not included in
the treatment of the Roy equations.

The way the fit to data goes is as follows.
The LASS data \cite {LASS} play a strong role in determining the
phase  of the amplitude.
The BES data determine the parameters of the strong $K_0(1430)$
observed there - considerably stronger than in LASS or E791
data.
The E791 data determine the magnitude of the amplitude,
which turns out to be accurately consistent with the phase variation
when the numerator of the production amplitude is taken to be constant.
This is a clear example how one can improve the precision of the
analysis by fitting several sets of data simultaneously.

Currently, the Belle, Babar and CLEO C collaborations make fits without
the Adler zero.
Fits including it are urgently needed to reduce the confusion.
Are they claiming their results demonstrate that Chiral Symmetry is
not broken in the $K\pi$ sector?
That seems unlikely in view of the fact that the Adler zero is very
well established in the heavier $\pi N$ sector.

\section {$f_0(980)$ and $a_0(980)$}
The whole question of how resonances can be attracted to thresholds
is discussed in a full length article \cite {Sync1} and will be
recapitulated here.
In the Breit-Wigner denominator of a resonance, there must be
dispersive terms in the real part
\begin {equation} m(s)_i =
\frac {1}{\pi} \rm {P} \int _{4m^2_i}
^\infty ds' \, \frac {g^2_i(s') \rho _i(s')} {s' - s}.
\end {equation}
In fact, the terms $ig^2_i\rho_i$ of Eq. (3) arise from the pole at
$s' = s$ in Eq. (4).

Fig. 2 shows $m_{KK}(s)$ and $g^2_{KK}\rho _{KK}(s)$ near the $KK$
threshold, using a form factor $e^{-3k^2}$, where $k$ is
centre of mass $KK$ momentum in GeV/c.
There is a cusp in $m_{KK}(s)$ at the threshold.
It is positive definite at the theshold, signifying additional
attraction there.
\begin{figure}[htb]
\begin{center} \vskip -8mm
\epsfig{file=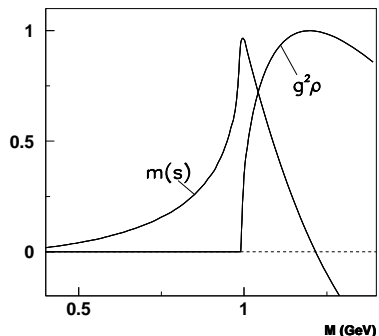,width=6.5cm}
\vskip -6mm
\caption {$m_{KK}(s)$ and $g^2_{KK}\rho _{KK}(s)$
for $f_0(980)$, normalised to 1 at the peak of $g^2_{KK}\rho _{KK}$.}
\end{center}
\end{figure}
If the cusp is superimposed on attraction from another
source, for example meson exchanges, a resonance can be generated by
the peak in $m_{KK}(s)$.
The locking of $a_0(980)$ and $f_0(980)$ to the $KK$ threshold is
explained naturally by this cusp mechanism.

Near threshold, the wave function of the resonance has a long tail,
like the deuteron.
This tail is purely mesonic; contributions from coloured quarks are
confined in the resonance at short range or in the decay mesons.
Confined quarks have kinetic
energy $k^2/2m = -(\hbar ^2/2m)\nabla ^2 \Psi$; here $k$ and $m$ are
quark momentum and effective mass and $\Psi$ is the wave function.
To create a bound state, extra potential energy must
compensate the kinetic energy of the confined particles.
This again pulls the resonance towards the threshold.
T\" ornqvist provides a formula allowing an estimate of the meson-meson
component \cite {Tornqvist}; for the $f_0(980)$ it is at least $60\%$
and for $a_0(980)$ at least $\sim 35\%$.

Other examples of resonances appearing at sharp thresholds are
known: $f_2(1565)$ at the $\omega \omega $ threshold,
$K_0(1430)$ at the $K\eta ' $ threshold, $\Lambda _C(2940)$ at the
$D^0p$ threshold, and $X(3872)$ at the $\bar D(1865)D^*(2007)$
threshold.

\section {How does Confinement work?}
The nonet of $\sigma$, $\kappa$, $a_0(980)$ and $f_0(980)$ may be
understood quantitatively in terms of meson exchanges, witness the
calculations from the Roy equations by Caprini et al. and
Descotes-Genon et al.
Janssen et al. have shown that the $f_0(980)$ may be fitted well in
terms of $K^*$ and $\rho$ exchanges in $u$ and $t$ channels
\cite {Janssen}.
However, most mesons and baryons are explained as quark-model
states.
There has been much debate about `molecules' and multi-quark states,
e.g. $qq\bar q \bar q$.

One of the mysteries of the quark model is how confined states
penetrate the confining potential and turn into mesons.
How, for example, does the $\rho$ `know' to have the right width
to create the $\sigma$ pole via the Roy equations?
My suggestion is that the confining potential arises as a cooperative
effect between QCD at short range (coloured quarks and gluons)
and meson exchanges at long range.
Meson exchanges are certainly present, witness the peaks observed
at small $t$ and $u$.
If both QCD and meson exchanges contribute, the well known eigenvalue
equation is
\begin {equation}
H \Psi  = \left(
\begin {array} {cc}
H_{11} & V \\
V & H_{22}
\end {array}
\right) \Psi = E\Psi,
\end{equation}
where $H_{11}$ and $H_{22}$ describe isolated $\bar qq$ and
meson-meson  configurations and $V$ describes mixing.
This mixing pushes the eigenstate down in mass.
This is the well known Variational Principle which minimises the
eigenvalue when states mix.
It is closely analogous to formation of a covalent bond in chemistry.

Van Beveren and Rupp \cite {eef} have proposed a similar idea.
They adopt a transition potential which couples states confined in a
harmonic oscillator potential to outgoing waves; they match the
logarithmic derivative at a $\delta$ function transition radius $\sim
0.65$ fm, though they mention the possibility
of a gradual transition with a $^3P_0$ dependence on radius.
The $\sigma$, $\kappa$, $a_0(980)$ and $f_0(980)$ emerge from the
continuum when the coupling constant to confined states increases.
The simplicity of their model makes it easy to follow
the movement of poles as this coupling constant is varied.
They illustrated results for the $\kappa$ and $a_o(980)$ poles.
The dependence of the $\sigma$, $\kappa$,
$a_0(980)$ and $f_0(980)$ poles on the coupling constant is
tabulated in Ref. \cite {joint}.
The model reproduces well the amplitudes for all these states
using a universal coupling constant and SU3 Clebsch-Gordan
coefficients.
The $a_0$ does not appear at the $\eta\pi$ threshold because of its
associated Adler zero.
If the coupling constant is increased by a factor 2.5,
the $\sigma$, $\kappa$ and $a_0$ all become bound
states.

There has been great enthusiasm for explaining states observed in
charmonium physics as 4-quark states.
However, Valcarce et al. \cite {Vijande} find from a detailed model
calculation that such states always appear higher in mass than
$c\bar c$ configurations with the same quantum numbers, unless
attractive interactions arise between diquarks.
At short range, coloured combinations could contribute to a lowering
of the energy of the $c\bar c n\bar n$ eigenstate.
However, the long-range tail of such an eigenstate is uncoloured,
taking us back to a meson-meson configuration.

Molecular states certainly exist as nuclei.
The nucleon-nucleon interaction is a classic example of meson
exchanges.
There is a repulsive core due to the Pauli principle
pushing identical quarks apart.

Oset, Oller and collaborators find that they can account for
many resonances from a chiral Lagrangian \cite {Oset} \cite {Oller}.
This is essentially a meson-meson interaction with Chiral
Symmetry Breaking built in.
They also include short-range $q\bar q$ interactions.
However, the short-range interactions need to be tuned to
reproduce accurately the masses, widths and branching ratios of
$f_0(1370/1310)$, $f_0(1500)$, $f_0(1710)$ and $f_0(1790)$.
In the $\pi N$ system, Donnachie and Hamilton showed in 1965
that meson exchanges are attractive for quantum numbers where
there are prominent resonances: $P_{33} (1232)$, $D_{13}(1520)$,
$D_{15}(1675)$, $F_{15}(1680$, etc. \cite {Hamilton}.

\section {$q\bar q$ states and Glueballs}
Ochs has waged a long campaign claiming
that Cern-Munich (CM) data \cite {Hyams} are better than the sum total
of all other data and disprove the existence of $f_0(1370)$ .
This is not true.
CM data give an excellent determination of $\pi \pi$ phase
shifts $\delta$ from 610 MeV to the $KK$ threshold (990 MeV).
However, above this there are strong correlations between $\delta$
and elasticity parameters $\eta$; consequently, errors on both become large.
These are illustrated in Figs. 4 and 6 of Au, Morgan and Pennington
\cite {AMP} and in Fig. 1 of Kaminski, Pelaez and Yndurain \cite {KPY};
errors for $\eta$ are very large because of lack of data determining
accurately the absolute magnitudes of $\pi \pi \to KK$, $\eta \eta$
and particularly the dominant channel $4\pi$.

It is easy to check these errors directly from errors on CM moments.
For every bin, four algebraic solutions emerge.
There is some constraint on the $\pi \pi$ D-wave from the line-shape of
$f_2(1270)$.
However, there is a substantial contribution from $f_2(1565)$, ignored
by Ochs.
It couples strongly to $\omega \omega$ \cite {Baker}; the
sub-threshold continuation of this channel is poorly known, as is its
coupling to $\rho \rho$.
Consequently, there is large uncertainty in the D-wave contribution to
moments $Y_4$, $Y_2$ and $Y_0$.
Also the $\rho(1450)$ and $\rho(1700)$ cannot be identified in CM data,
leading to further poorly identified contributions to $Y_2$, $Y_1$ and
$Y_0$.
All these systematic uncertainties accumulate in $Y_0$, which
determines the S-wave intensity; it can be fitted very flexibly.
Fig. 3 shows my fit to CM moments using fully analytic functions (a
major constraint).
Some mixing is required (and expected) between $f_0(1370)$, $\sigma$
and $f_0(1500)$, which overlap strongly. Mixing induces a
phase rotation of the $f_0(1370)$, see \cite {f01370}.

\begin {figure}  [htb]
\begin {center}
\vskip -12mm
\epsfig{file=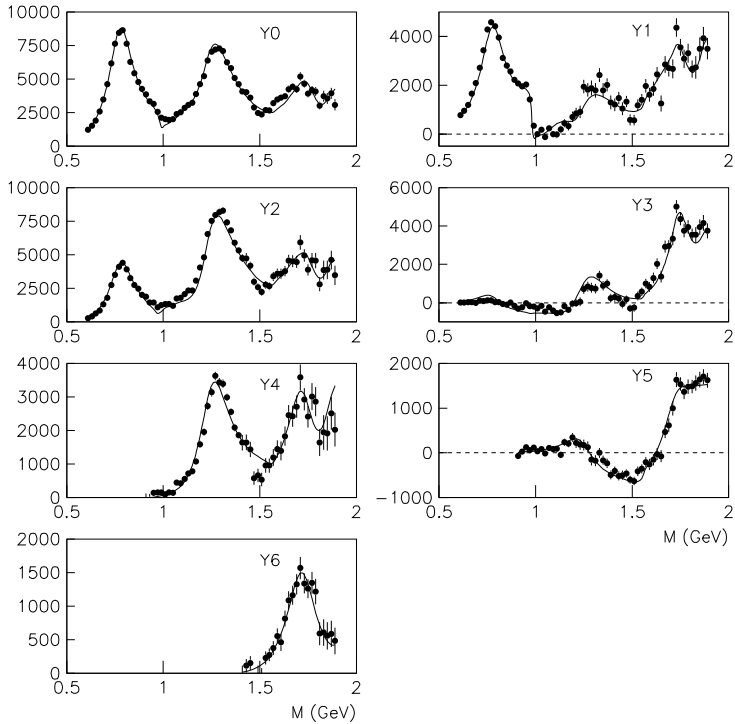,width=11cm}\
\vskip -8mm
\caption{My fit to Cern-Munich moments with $Y_0 - Y_6$}
\end {center}
\end {figure}

The problem in identifying $f_0(1370)$ is that its $2\pi$ decay
is obscured in many sets of data by $f_2(1270)$.
However, it produces a clearly visible peak in $\bar pp \to
\eta \eta \pi^0$ at rest; this peak cannot be due to
$f_2(1270)$ which has a branching fraction to $\eta \eta$ of only
0.4\%.
It is also visible as a peak in BES data for $J/\Psi \to
\phi\pi^+\pi^-$ \cite {phipp}, where angular correlations between
production and decay of the $\phi$ separate $J^P = 0^+$ and $2^+$
cleanly. It was first observed in 3 sets of data on $\pi \pi \to KK$
from ANL and BNL \cite {Cohen} \cite {Polychronakos} \cite {Longacre}.
Its parameters are best determined by Crystal Barrel data on $\bar pp
\to 3\pi ^0$ at rest in liquid and gaseous hydrogen; these data
separate $^1S_0$ and $^3P_1$ annihilation, which give independent
determinations of mass and width agreeing within $\pm 5$ and $\pm 10$
MeV respectively.
Crystal Barrel data on charge combinations of $\bar
pp \to KK\pi$ give consistent parameters with all other data \cite
{Andrei}.

The resonance mass is $1309 \pm 15$ MeV with a full-width at
half maximum of $207 \pm 15$ MeV.
However, the rapid opening of the $4\pi$ threshold moves
the peak in $4\pi$ higher by $\sim 80$ MeV because of phase space.
The opening of the $4\pi$ channel was not taken into account
in most analyses, leading to confusion and inflated errors.
The $f_0(1370)$ was observed by Gasparo \cite {Gaspero} and then
by Crystal Barrel, Obelix and WA102 in many sets of $4\pi$ data
\cite {PDG}.
When one allows for $4\pi$ phase space and the associated dispersive
effect, all observations where both $f_0(1310/1370)$ and $f_0(1500)$
have been fitted are close to consistency within errors.

With the $f_0(1370)$, $f_0(1500)$ and $f_0(1710)$ well established,
there is one state too many to fit as $\bar nn$ and $s\bar s$.
The glueball predicted in this mass range by Lattice Gauge
calculations mixes with the $q\bar q$ states.
The $f_0(1710)$ decays almost purely to $KK$, with a possible small
$\eta \eta$ decay.
The $f_0(1500)$ behaves as an octet state with $\Gamma
(\eta \eta)/\Gamma (\pi \pi) = 0.145 \pm 0.027$ and $\Gamma (K\bar
K)/\Gamma (\pi \pi) = 0.246 \pm 0.026$.

An intriguing result is the sharp $\omega \phi$ signal reported by
the BES collaboration \cite {omegaphi}, peaking at 1812 MeV, just
above the $\phi \omega$ threshold at 1801 MeV.
A purely threshold effect should peak much higher.
It is therefore probably due to $f_0(1790)$,
a resonance clearly separated from $f_0(1710)$ in BES data on
$J/\Psi \to \omega KK$ \cite {WKK}, $\omega \pi \pi$ \cite {BESWPP},
$\phi \pi \pi$ and $\phi KK$ \cite {phipp}.
There is a strong $f_0(1710)$ peak in
$\omega KK$ data, but nothing in $\omega \pi \pi$ despite large
statistics.
Conversely, the $f_0(1790)$ appears clearly in $\phi \pi
\pi$, but not in $\phi KK$.
There is a factor 22 difference in decay
branching ratios to $\pi \pi$ and $KK$, requiring separate $f_0(1710)$
and $f_0(1790)$.
The $f_0(1790)$ was first observed in $J/\Psi \to
\gamma 4\pi$ \cite {Mark3} \cite {g4pi}. It makes a natural radial
excitation of $f_0(1370)$.

The $\phi \omega$ decay can arise naturally from a glueball component
in $f_0(1790)$ \cite {Bicudo}.
A glueball is a flavour singlet with flavour content
\begin {equation}
F = (u\bar u + d\bar d + s\bar s)(u\bar u + d\bar d + s\bar s).
\end {equation}
[The $s\bar s$ component might be enhanced with respect to $n\bar n$
by a factor $f^2_K/f^2_\pi = 1.21$.]
The component $2(u\bar u + d\bar d)s\bar s$ can make $4\omega \phi$
or $2(K^{*0}\bar K^{*0} + K^{*+}K^{*-})$ or a linear combination.
BES I data on $J/\Psi \to \gamma (K^+\pi ^- K^-\pi ^+)$  show that
the $\gamma (K^*\bar K^*)$ channel contains no significant $0^+$
signal \cite {KstKst}.
A signal with the same magnitude as that of
$J/\Psi \to \gamma (\omega \phi)$ would be
conspicuous near 1800 MeV.
Its absence may be qualitatively explained by the larger
phase space for $(u\bar u + d\bar d)s\bar s$ in $KK$ decays than
in $K^* \bar K^*$.

\section {Conclusions}
In fitting $\sigma$ and $\kappa$, it is essential to include the
Adler zero explicitly into the Breit-Wigner denominator - or
demonstrate that such a fit is unacceptably bad.
The Adler zero needs to be included also into the $K_0(1430)$.
It would be good if competing experimental groups would join forces,
like LEP groups, to try and achieve an optimum compromise between
different sets of data and iron out inconsistencies.
Experience is that different sets of data illuminate different
aspects required for an optimum fit.

The $\sigma$, $\kappa$, $a_0(980)$ and $f_0(980)$ fit naturally
according to meson exchanges. Why look further?
The cusp at the $KK$ threshold plays an
essential role in locking $a_0$ and $f_0$ to this threshold.
The cusp mechanism can attract a resonance
over a surprisingly large mass interval of $\sim \pm 100$ MeV.
At the threshold, zero point energy is minimised by the long-range
tail of the wave function.
Mixing between quark configurations and meson-meson states minimises
the energy of the linear combination in a way analogous to the
formation of a covalent bond in chemistry.
This idea is similar to the idea of Duality between resonances
and particle exchanges.

\section {Acknowledgements}
I wish to thank Prof. G. Rupp and Prof. E. van Beveren for
extensive discussions and for their efforts in organising a workshop
on Scalar Mesons to mark the 70th birthday of Mike Scadron.
I am very grateful to Mike Scadron for illuminating the role of
PCAC and the linear $\sigma$ model on many occasions.

\end {document}